\documentclass[10pt,twocolumn]{article}

\usepackage[utf8]{inputenc}
\usepackage[T1]{fontenc}
\usepackage{amsmath,amssymb}
\usepackage{graphicx}
\usepackage{booktabs}
\usepackage{hyperref}
\usepackage{listings}
\usepackage{xcolor}
\usepackage[margin=0.85in]{geometry}
\usepackage{enumitem}
\usepackage{caption}
\usepackage{algorithm}
\usepackage{algpseudocode}
\usepackage{multirow}
\usepackage{tabularx}
\usepackage{fancyvrb}
\usepackage{tikz}
\usetikzlibrary{positioning,arrows.meta,fit,backgrounds}

\title{\textbf{Elastic Gang: Per-Token Membership Change for a\\Hard-Barriered LLM Inference Gang Co-Scheduled with OS Processes}}

\author{
  Daeyeon Son\\
  Independent Researcher\\
  Republic of Korea\\
  \texttt{sdy1350@gmail.com}
}

\date{July 2026 --- draft}

\begin{document}
\maketitle

\begin{abstract}
On-device LLM decoding is a hard-barriered CPU-SIMD computation that
wants every core for milliseconds per token --- while the rest of the
OS wants those same cores continuously. A barriered gang cannot simply
be thrown into a preemptive scheduler: an unannounced departure
deadlocks a barrier, and an unannounced arrival silently corrupts
logits. I present the \emph{elastic gang} of Anima~OS, a bare-metal
\texttt{x86\_64} Rust kernel in which the inference gang is a
first-class schedulable entity whose core membership may change
\emph{between any two tokens}. The core mechanism is an
\textbf{ACK-latched epoch protocol that never waits on a named core}
--- a seqlock-style generation-tagged participant latch composed with
RCU/epoch-style membership consent: each token's participant set is
the intersection of the cores the gang \emph{requested} and the cores
that \emph{acked the current epoch}, snapshotted into a single tagged
latch word; an un-acked core is simply outside \emph{this} token and
joins at most one token later, while work-stealing absorbs its rows.
Displaced general processes migrate and keep running; cores return to
them the moment a generation ends. On a real AMD Zen~5 machine
(8\,C/16\,T), inference output is \emph{bit-exact} under verified
per-token membership change on both a 135M and a 7B model --- the
property that makes elasticity safe in a kernel whose safety gate
reads logits.

Against fair static core partitions (93--100\% of both solo
references; the static baseline is a restriction of the same shipped
binary, making the sweep an ablation of policy alone), elastic
membership Pareto-dominates: at intermediate inference duty cycles it
delivers 1.75$\times$ (25\%), 1.52$\times$ (50\%), and 1.28$\times$
(75\%) the general throughput of a static 8-core split at equal or
better inference throughput, recovers all eight stranded cores when
inference is idle, and converges to the split at saturation ---
tracking a frontier no fixed partition tracks at more than one point.
Returning a lent core costs $0.22\,\mu$s (p50); acquiring a busy,
tenant-occupied core costs one scheduling quantum ($\approx$16\,ms)
by design --- a running tenant is never preempted mid-slice ---
amortized over the generation, while the per-token participation
latch is a single tagged atomic read. Decode throughput saturates at
gang width 8 in two different memory regimes (DRAM-bound 7B at
12.7\,tok/s; cache-resident 135M, which \emph{loses} throughput past
the knee), so ceding cores past the knee is nearly free ---
elasticity auto-sizes the gang to the knee online.
\end{abstract}

\section{Introduction}
\label{sec:intro}

An operating system that hosts LLM-driven agents runs two populations
of work with opposite shapes. \emph{General processes} --- shells,
services, agent runtimes, I/O --- are preemptible, independent
threads: the textbook workload of a weighted round-robin or
virtual-time scheduler. \emph{LLM inference} is the opposite: a
single-program-multiple-data computation in which every participating
core executes the same operator stream ($\sim$226 barriers per token
in my engine), synchronizing at hard barriers between matrix-vector
waves. During a token, the gang wants every core it can get; between
generations it wants none.

Committing cores to either population strands the other's cycles. A
static ``inference partition'' of $K$ cores idles whenever no
generation is in flight --- and on-device agent inference is bursty by
construction (my agents' generations are 32-token syscall-bracketed
bursts; interactive \texttt{ask}-style jobs arrive at human cadence).
A static \emph{general} partition throttles every generation to
$16-K$ cores. And a barriered gang cannot simply be thrown into the
preemptive scheduler: preempting one member mid-token stalls all
members at the next barrier, and naive membership changes either
deadlock a barrier on a departed core or silently corrupt the output
when a core walks a token it was not latched into
(\S\ref{sec:problem}).

This paper reports the design, implementation, and silicon measurement
of the \textbf{elastic gang}: the inference gang as a first-class
kernel-scheduled entity that \emph{borrows} cores from the general
population when a generation begins, changes membership per token, and
\emph{returns} the cores --- work-conservingly --- the moment the
generation ends. The enabling observation is that a hard-barriered
computation can tolerate elastic membership if, and only if, no
barrier ever waits on a \emph{named} core. My protocol achieves this
with two pieces:

\begin{itemize}[leftmargin=1.2em,itemsep=1pt]
\item an \textbf{ACK-latched epoch protocol}: the gang bumps a global
  epoch at generation start \emph{and} end; a lending core
  acknowledges the current epoch after non-destructively parking its
  tenant; a core participates in a token iff its ack matches the
  current epoch at latch time (\emph{requested} $\cap$ \emph{acked});
\item a \textbf{generation-tagged participant latch}: the participant
  set is snapshotted once per token into a single packed word
  (\texttt{gen}$_{lo32} \ll 32 \mid$ \texttt{mask}) published by the
  token-start bump; every barrier wait site and every worker's
  self-exclusion check reads \emph{this one word}, so the wait-side
  and worker-side views are provably the same snapshot, and any
  membership change takes effect only at the next token.
\end{itemize}

A core that has not (yet) acked is simply not in this token ---
nothing waits for it; the engine's work-stealing absorbs its rows; it
joins at the next token boundary. A core that departs can never be
counted again, because both generation start and end bump the epoch,
invalidating every outstanding ack by construction. Displaced general
processes are migrated off lent cores by the scheduler (paced, two per
scheduler pass) and keep running on the survivor cores.

I measure the system on real hardware (AMD Ryzen~9800X3D, Zen~5,
8~cores/16~hardware threads, DDR5-6000), on two operating points ---
SmolLM2-135M ($\sim$0.75\,ms/token, stressing per-token elasticity)
and Qwen2.5-7B ($\sim$79\,ms/token, bandwidth-bound) --- against
static-partition baselines that pass explicit fairness gates
(93--100\% of both solo references, \S\ref{sec:eval-sanity}).

\paragraph{Contributions.}
\begin{enumerate}[leftmargin=1.4em,itemsep=1pt]
\item \textbf{The never-wait-on-a-named-core ACK-latch protocol
  (\S\ref{sec:design}).} Per-token gang membership change for a
  hard-barriered SIMD computation --- a seqlock-style
  generation-tagged participant latch composed with RCU/epoch-style
  membership consent --- with the invariant argument for why no
  barrier can hang on a departed or late core and why no stale ack
  can ever be latched.
\item \textbf{Bit-exactness under membership churn, on silicon
  (\S\ref{sec:eval-m4}).} Under verified membership change (the
  participant mask really changed, through six distinct sets) the
  greedy token sequence and the final logits are byte-identical to a
  fixed-width run, on both the 135M and the 7B model, with a
  negative control proving the equality test can fail. A companion
  experiment pins the gang width entirely and lets \emph{real
  lending} drive per-token membership; the output is again
  byte-identical, so the invariance holds under both membership
  drivers the protocol has. In this
  kernel a lost or extra gang core is a \emph{correctness and
  safety} event, not a performance event: logits feed the safety
  gate~\cite{animaos-mcp}.
\item \textbf{Measured work-conservation Pareto dominance
  (\S\ref{sec:eval-m3}).} Because the static-$K$ baseline is a
  restriction of the same shipped binary, the duty-cycle sweep is an
  ablation of scheduling policy alone. At intermediate duty cycles
  --- the operating region a deployment actually lives in ---
  elastic membership delivers 1.75$\times$/1.52$\times$/1.28$\times$
  the general throughput of a static 8-core split (duty
  25/50/75\%) at equal-or-better inference throughput, recovers
  all eight stranded cores at idle, converges to the split at
  saturation, and dominates a static 12-core split's general
  throughput 2.3--4.7$\times$ everywhere. No fixed $K$ tracks the
  frontier; the elastic gang auto-sizes online.
\item \textbf{Auto-sizing to the saturation knee, in two memory
  regimes (\S\ref{sec:eval-m7}).} Decode throughput saturates at
  gang width 8 both for the DRAM-bandwidth-bound 7B model
  (12.7\,tok/s) and for the cache-resident 135M model, which
  \emph{declines} past the knee; lending cores away from the gang
  is therefore nearly free for inference, which is what makes
  elasticity profitable rather than merely possible.
\end{enumerate}

The mechanism already carries per-tenant trust state, and
\S\ref{sec:eval-c32} measures the consequence: an agent's
kernel-maintained constitutional trust score, wired in as a
\emph{lend-admission quota}, monotonically dials the machine
between the elastic sweep's two endpoints --- governance as a
first-class scheduler lever. I report it as a measured property of
the substrate rather than as a separate contribution.

\section{Background and Problem}
\label{sec:problem}

\subsection{Setting: one kernel, two populations}
\label{sec:setting}

Anima~OS is a bare-metal \texttt{x86\_64} kernel written in
\texttt{no\_std} Rust ($\sim$232{,}000 lines; $\sim$290{,}000
across the project including hosted tooling) whose distinguishing
subsystem is an in-kernel LLM inference engine: a GGUF loader,
AVX-512 quantized kernels, and a graph executor that runs one token
as a fixed operator stream (matvec waves separated by $\sim$226
barriers) across a set of worker cores with row-range work-stealing
inside each wave~\cite{animaos-probelogits}. The engine backs both
interactive jobs (\texttt{ask}, document AI) and agent syscalls
(\texttt{llm\_generate}), and is the substrate of the kernel's
safety layer~\cite{animaos-mcp}, so inference is not an optional
accessory --- it is resident OS work.

The same kernel schedules \emph{general processes}: preemptive
tenants under a weighted-round-robin driver with integer virtual
time (per-process \texttt{delivered} counters), per-core run queues,
a dynamic slot table, and a CAS-based migration primitive
(\texttt{RUNNABLE}$\to$\texttt{MIGRATING}). There is no fixed
``inference core'' role: any core can be a barrier worker in one
millisecond and a tenant host the next.

Two facts shape the design space:

\textbf{(F1) Decode saturates well below machine width.}
Generating one token streams essentially the whole quantized
weight set through the memory hierarchy. For serving-class models
the bound is DRAM: on DDR5-6000 (dual channel, $\sim$75\,GB/s), a
4.2\,GB Q4\_0 7B model has a $\sim$55\,ms/token bandwidth floor; I
measure 78.6\,ms/token at width 8, and adding cores beyond 8 buys
$<$1\% (\S\ref{sec:eval-m7}). (A cache-resident 135M model
saturates at the same width for a different reason ---
\S\ref{sec:eval-m7}.) Compute-side optimization of the gang is
exhausted; the remaining lever is \emph{what the other cores do}.

\textbf{(F2) Inference is bursty.} Agent generations are short
(32-token bracketed bursts by construction); interactive generations
arrive at human cadence; safety classifications are single
prefill-plus-one-logit probes. Between bursts, a dedicated
inference set is pure waste.

\subsection{Why a barriered gang is not a bag of threads}
\label{sec:notfungible}

A reader steeped in $\mu$s-scale core reallocation --- Shenango
and Caladan~\cite{shenango,caladan} --- may ask why the gang
cannot simply grow and shrink like their batch workloads. The
answer is that their reallocated unit is a \emph{fungible thread}:
any worker can serve any request, and a revoked core's work is
completed by any other. A barriered SIMD gang is not fungible in
this sense; naive membership change fails in exactly three ways:

\begin{enumerate}[leftmargin=1.4em,itemsep=1pt]
\item \textbf{Barrier deadlock on departure.} Barrier waits are
  counted against a participant set. If a named core leaves after
  the wait-side computed its set but before arriving at the
  barrier, the remaining members wait forever. There is no
  ``un-latch'': once a latch has counted a core for a token, that
  core is \emph{obligated} to walk it.
\item \textbf{Silent corruption on late or double entry.} If a core
  joins and walks a token it was not latched into --- e.g.\ its
  read of the participant set straddled the token-start bump and it
  saw the \emph{next} token's mask --- it steals rows concurrently
  with the token's real participants. Nothing crashes; the logits
  are silently wrong. In a kernel whose safety gate \emph{reads
  logits}~\cite{animaos-mcp}, a silently wrong forward pass is a
  security event, not a performance bug.
\item \textbf{Stale-ack resurrection.} A consent flag that survives
  a generation boundary can count a core that has long since
  returned to running tenants --- the same hang as (1), one
  generation later.
\end{enumerate}

Failure modes (1)--(3) are why classic gang
scheduling~\cite{ousterhout82,feitelson92} co-schedules \emph{whole
time slices} of fixed membership, and why the demand-based and
flexible co-scheduling lines~\cite{sobalvarro95,fcs03,wiseman03}
relax \emph{which jobs} run together but not \emph{membership of a
running barrier group}. The elastic gang's job is to make
membership change safe at the finest granularity the computation
admits: the token boundary.

\subsection{Design goals}
\label{sec:goals}

\begin{itemize}[leftmargin=1.2em,itemsep=1pt]
\item[\textbf{G1}] \emph{Work conservation.} No core idles while
  either population has runnable work, in either direction.
\item[\textbf{G2}] \emph{Never wait on a named core.} No barrier,
  and no phase of the protocol, may block on a specific core
  acting; membership is whoever showed up by the latch.
\item[\textbf{G3}] \emph{Bit-exactness.} Output must be
  byte-identical to a fixed-membership run under any membership
  schedule the protocol permits.
\item[\textbf{G4}] \emph{Displaced work survives.} A
  \emph{migratable} tenant on a lent core is parked
  non-destructively, migrated, and keeps running; lending must
  never kill or wedge a process. (A tenant pinned to a lendable
  core waits out the loan --- a disclosed worst case measured in
  \S\ref{sec:eval-m3}.)
\item[\textbf{G5}] \emph{Zero cost when disabled.} With no lendable
  core designated, the entire path must be provably dead
  (bit-identical kernel behavior --- enforced by my regression
  gate).
\end{itemize}

\section{Design}
\label{sec:design}

\subsection{Overview}
\label{sec:overview}

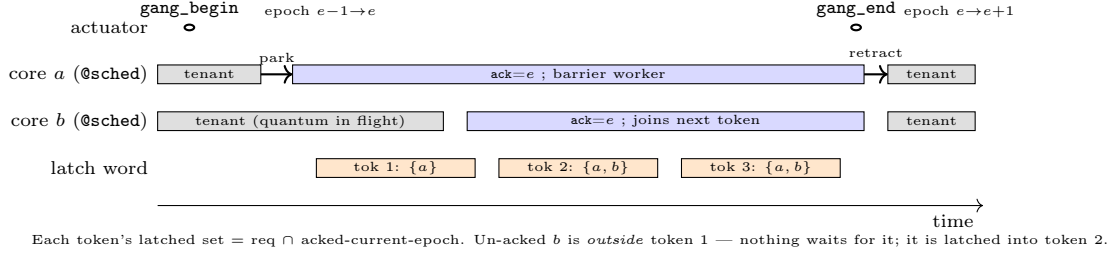
\begin{figure*}[t]
\centering
\begin{tikzpicture}[font=\scriptsize, x=1.05cm, y=0.62cm]
  \node[anchor=east] at (0,4) {actuator};
  \node[anchor=east] at (0,3) {core $a$ (\texttt{@sched})};
  \node[anchor=east] at (0,2) {core $b$ (\texttt{@sched})};
  \node[anchor=east] at (0,1) {latch word};
  \draw[->] (0,0.2) -- (10.4,0.2) node[below left]{time};
  \draw[thick] (0.4,4) node[above]{\texttt{gang\_begin}} circle (0.06);
  \node[above] at (2.05,4.02) {\tiny epoch $e{-}1{\to}e$};
  \draw[thick] (8.8,4) node[above]{\texttt{gang\_end}} circle (0.06);
  \node[above] at (10.1,4.02) {\tiny epoch $e{\to}e{+}1$};
  \draw[fill=gray!25] (0,2.8) rectangle (1.3,3.2);
  \node at (0.65,3) {\tiny tenant};
  \draw[thick,->] (1.3,3) -- (1.7,3) node[midway,above]{\tiny park};
  \draw[fill=blue!15] (1.7,2.8) rectangle (8.9,3.2);
  \node at (5.3,3) {\tiny \texttt{ack}=$e$ ; barrier worker};
  \draw[fill=gray!25] (9.2,2.8) rectangle (10.3,3.2);
  \node at (9.75,3) {\tiny tenant};
  \draw[thick,->] (8.9,3) -- (9.2,3);
  \node at (9.05,3.45) {\tiny retract};
  \draw[fill=gray!25] (0,1.8) rectangle (3.6,2.2);
  \node at (1.8,2) {\tiny tenant (quantum in flight)};
  \draw[fill=blue!15] (3.9,1.8) rectangle (8.9,2.2);
  \node at (6.4,2) {\tiny \texttt{ack}=$e$ ; joins next token};
  \draw[fill=gray!25] (9.2,1.8) rectangle (10.3,2.2);
  \node at (9.75,2) {\tiny tenant};
  \draw[fill=orange!20] (2.0,0.8) rectangle (4.0,1.2);
  \node at (3.0,1.0) {\tiny tok 1: $\{a\}$};
  \draw[fill=orange!20] (4.3,0.8) rectangle (6.3,1.2);
  \node at (5.3,1.0) {\tiny tok 2: $\{a,b\}$};
  \draw[fill=orange!20] (6.6,0.8) rectangle (8.6,1.2);
  \node at (7.6,1.0) {\tiny tok 3: $\{a,b\}$};
  \node[align=left] at (5.2,-0.55) {\tiny Each token's latched set $=$ req $\cap$ acked-current-epoch. Un-acked $b$ is \emph{outside} token 1 --- nothing waits for it; it is latched into token 2.};
\end{tikzpicture}
\caption{One generation under the ACK-latched epoch protocol. The
participant set of each token is \emph{requested} $\cap$
\emph{acked-current-epoch}, snapshotted into one gen-tagged word at
token start. No barrier ever waits on a named core: a late core is
simply outside this token; a departed core can never be counted
again because both \texttt{gang\_begin} and \texttt{gang\_end} bump
the epoch.}
\label{fig:protocol}
\end{figure*}

Figure~\ref{fig:protocol} shows one generation. Cores designated
\texttt{@sched} run a resident scheduler loop serving general
tenants. When a generation begins, the kernel (the \emph{actuator}
is the coordinating core, never a barrier participant mid-wait)
executes \texttt{gang\_begin}: claim ownership, bump the epoch,
raise the active flag. Each lend-eligible \texttt{@sched} core
notices the active gang at its loop top, parks its current tenant
non-destructively, and \emph{acks the current epoch}. From then on
it behaves exactly like a native inference worker --- walking
tokens, waiting at barriers, MWAIT-parking between tokens --- until
the epoch changes, at which point it retracts its ack and resumes
its scheduler loop. Tenants stranded on lent cores are migrated by
the scheduler's service pass (\S\ref{sec:lending}); when
\texttt{gang\_end} bumps the epoch and rings the dispatch doorbell,
the lent cores wake, retract, and their queues resume.

\subsection{The ACK-latched epoch protocol}
\label{sec:ack}

In synchronization-primitive terms, the protocol composes two
well-known shapes into a non-obvious whole: membership consent is
RCU/epoch-style (a reader-visible epoch that both start and end
advance, so stale consent dies by construction), and per-token
membership publication is seqlock-style (a single generation-tagged
word whose tag lets every reader detect and reject a straddled
read). Four pieces of state, all lock-free atomics with sequentially
consistent edges on the protocol boundary:

\begin{itemize}[leftmargin=1.2em,itemsep=1pt]
\item \texttt{GANG\_EPOCH}: a global generation-ownership epoch,
  \textbf{initialized to 1} and bumped at generation start
  \emph{and} end. Starting at 1 (not 0) means the all-zero boot
  state of the ack array can never equal the epoch --- a
  $0{=}0$ match would latch a core that never parked its tenant.
\item \texttt{GANG\_ACK[$i$]}: per-core (cache-line padded) --- the
  epoch core $i$ acked; 0 means ``not lending.'' Core $i$
  participates in a token iff
  $\texttt{GANG\_ACK}[i] = \texttt{GANG\_EPOCH}$ \emph{at latch
  time}. This comparison has exactly one reader site
  (\S\ref{sec:latch}).
\item \texttt{GANG\_ACTIVE}: the lend invitation flag.
\item \texttt{GANG\_OWNER}: single-owner CAS
  ($0\!\to\!\textsc{owner}$), \S\ref{sec:owner}.
\end{itemize}

\texttt{gang\_begin(owner)} performs, in order: owner CAS (failure
$\Rightarrow$ the caller runs \emph{un-ganged} at its base width ---
it must never abort a generation); epoch bump; active-flag set. If
no \texttt{@sched} core exists it returns false before touching
anything --- goal G5's provably-dead path.

\texttt{gang\_end(owner)} is owner-keyed and idempotent: epoch bump
(no stale ack can ever match again --- the departure guarantee),
clear active, release owner, \emph{then} ring the dispatch doorbell
so lent cores wake off MWAIT and their parked tenants resume. It is
called at all four generation exit paths (done, faulted, abandoned,
lease-fault-cleared).

A lending core's service loop (\texttt{gang\_service}) has one
ordering rule that is load-bearing:
\textbf{resync-before-ack}. The core first snapshots the engine's
token generation counter ($g_{last}$), and only then stores its ack.
Ack-first would allow a token latch to count this core for a token
whose generation number the later resync then skips ---
latched-but-never-walks, a barrier hang. With resync-first, any
token latched against our ack has its generation bump strictly
after our ack, hence after our resync, hence $g > g_{last}$
$\Rightarrow$ it is walked. Symmetrically, the ack is \emph{never}
re-synced or re-issued mid-epoch: a re-resync while the ack is
latch-visible could jump $g_{last}$ forward over a token the latch
already obligated us to walk. The ack simply stays up for the
epoch's lifetime; departure happens only on epoch change, which is
safe by construction (no latch computed after the bump can count
the stale ack).

\subsection{The generation-tagged participant latch}
\label{sec:latch}

Membership must change only at token boundaries (G3), and the
barrier wait sites and the workers' self-exclusion checks must see
the \emph{same} set (failure mode 2 of \S\ref{sec:notfungible}).
Both are enforced by funneling the entire protocol through one
word:

\begin{lstlisting}[caption={The single latch-computation site
(abridged from \texttt{graph.rs}). \texttt{GANG\_ACK}/\texttt{GANG\_EPOCH}
are read \emph{only} here --- never live at a barrier wait site or
inside a worker.},label={lst:latch}]
// packed: (token_gen as u32) << 32 | mask
fn participant_mask(n_aps: usize) -> u32 {
  let mut m = 0u32;
  for i in 0..n_aps.min(32) {
    let excluded = is_util_core(i)
      || is_width_reserved(i)
      || (is_sched_ap(i)
          && !gang_acked_current(i)); // ACK-latch
    if !excluded { m |= 1 << i; }
  }
  m
}
// published by the token-start Release bump;
// all 5 barrier wait sites + the worker
// self-exclude read THIS one tagged word.
\end{lstlisting}

The mask is computed once per token, packed with the low 32 bits of
the token's generation number, and stored immediately before the
Release increment of the generation counter that starts the token
--- so the bump publishes the latch. All six consumers (five
barrier wait sites, plus each worker's self-exclusion on entry)
read this one word: wait sites take the low half; a worker compares
the high-half tag against the generation it is servicing and treats
a mismatch as \emph{excluded}. The tag closes failure mode 2: a
worker whose read straddles a token-start bump sees a tag that is
not the token it was admitted to, and bails instead of
double-walking. With no gang in flight the tag always matches and
the mask is bit-identical to the pre-elastic kernel --- G5 again.

\paragraph{The never-wait property.} Putting
\S\ref{sec:ack}--\ref{sec:latch} together: (i) a core that has not
acked by latch time is not in the mask --- no wait site counts it,
no row is reserved for it; work-stealing partitions the wave over
whoever \emph{is} in the mask; (ii) a core that acked is obligated
only for tokens latched while its ack matches the epoch, and the
resync-before-ack order guarantees it walks all of them; (iii) a
departed core's ack is dead the instant the epoch bumps, and the
last token latched before the bump was walked by (ii) before the
core's service loop observed the change and retracted. At no point
does any component wait for a \emph{particular} core to ack,
arrive, or leave. Membership latency is therefore bounded by one
token, and barrier integrity is independent of membership timing.

\subsection{Lending, migration, and return}
\label{sec:lending}

\textbf{Which cores lend.} Only cores inside the engine's
configured worker range are lend-eligible (a core outside the
32-bit latch range must never be latched); a designated
high-priority scheduler core never lends. Eligibility is evaluated
at the resident loop top --- lending is always the lent core's own
act (self-claiming), never a remote grab.

\textbf{What happens to the tenant.} The resident loop parks its
current tenant non-destructively (state and FPU saved to its slot;
the slot stays \texttt{RUNNABLE}) before acking. The scheduler's
per-pass service then migrates parked tenants off lent cores using
the CAS migration primitive (\texttt{RUNNABLE}$\to$%
\texttt{MIGRATING}$\to$re-enqueue), deliberately paced at
$\le$2 migrations per pass to bound service-pass latency; a
companion rebalance pass smooths queue skew afterwards. Between
tokens, a lent core also reaps its own kill-flagged parked tenants,
so process termination is never deferred behind a whole generation.

\textbf{Return.} \texttt{gang\_end}'s doorbell wakes every lent
core out of MWAIT; each observes the epoch change, retracts its ack
(sub-microsecond, \S\ref{sec:eval-m1}), and falls back into its
scheduler loop, picking up whatever its queue holds. Return is
work-conserving by construction: there is no draining phase and no
handshake with the actuator.

\subsection{Ownership and fault containment}
\label{sec:owner}

Two generations must never both drive the gang: generation A's end
would strip generation B's lent cores mid-forward. The owner CAS
serializes this --- a loser simply runs un-ganged at base width
(never aborts), and only the matching owner's \texttt{gang\_end}
releases. Fault paths are the sharp edge: an agent's generation
runs inside a syscall, and a page fault there long-jumps out with
no stack unwind, skipping the bracket's \texttt{gang\_end} --- which
would park every lent core's tenants forever. The kernel therefore
maintains a per-CPU \emph{bracket-open} flag set before
\texttt{gang\_begin} and cleared after \texttt{gang\_end}; the page
fault handler's agent-context branch consumes the flag before
either long-jump route and releases the agent-owned gang. The
interactive job path is covered by its own watchdog deadline. Both
paths funnel into the same idempotent, owner-keyed
\texttt{gang\_end}.

\subsection{Invariant summary}
\label{sec:invariants}

\begin{table}[h]
\centering\scriptsize
\begin{tabular}{@{}p{0.6cm}p{6.6cm}@{}}
\toprule
\textbf{I1} & Membership changes only at token boundaries (latch snapshot published by the token-start bump). \\
\textbf{I2} & No barrier waits on a core absent from the current token's latch (mask $=$ requested $\cap$ acked-current). \\
\textbf{I3} & No stale ack is ever latched (epoch bumped at start \emph{and} end; epoch starts at 1 vs.\ zero-initialized acks). \\
\textbf{I4} & Wait-side and worker-side read the same snapshot (single latch word; single \texttt{GANG\_ACK} reader site; gen tag rejects straddled reads). \\
\textbf{I5} & An acked core walks every token latched against its ack (resync-before-ack; no mid-epoch re-resync). \\
\textbf{I6} & Exactly one generation owns the gang (owner CAS; losers run un-ganged; owner-keyed release). \\
\textbf{I7} & Every generation exit path --- including \#PF long-jumps --- releases the gang (bracket flag consumed by the fault handler; watchdog on the interactive path). \\
\textbf{I8} & With no lendable core designated, the whole path is dead code (bit-identical kernel; regression-gated). \\
\bottomrule
\end{tabular}
\caption{Elastic-gang protocol invariants.}
\label{tab:invariants}
\end{table}

\section{Implementation}
\label{sec:impl}

\subsection{Kernel substrate}

The mechanism lives in the SMP layer and the graph executor of
Anima~OS's $\sim$232{,}000-line \texttt{no\_std} bare-metal kernel
(the inference engine, preemptive scheduler, and governance layers
are described in the companion
papers~\cite{animaos-probelogits,animaos-mcp}). All protocol state
is fixed-size static atomics; the per-core ack array is cache-line
padded (a false-sharing audit of the barrier path preceded this
work). Protocol edges are \texttt{SeqCst}: the ack/epoch/latch
interplay is a Dekker-style publication pattern, and the two
silicon-only failure modes found during development (a stale-ack
departure race and a latch read straddling the token bump) both
motivated strengthening, not weakening, the ordering. Lent cores
park on MWAIT armed on a dispatch doorbell; every writer a lent
core polls rings the doorbell after its store (a missed ring is a
silicon hang that virtualized testing does not model --- KVM lacks
MONITOR/MWAIT --- so this discipline is verified on hardware).

\subsection{The static-partition baseline as a mode}
\label{sec:staticmode}

The comparison target of \S\ref{sec:eval-m3} is a \emph{static}
$K$-core inference partition. Crucially, this is a
\emph{restriction} of the shipped mechanism, not separate
machinery: in static mode, \texttt{gang\_begin} returns false
before touching any state, which is bit-identical to the
no-lendable-core dead path (I8), and general-process placement
hard-excludes the $K$ inference cores (in elastic mode they are
merely handicapped). Inference then runs at fixed width $K$ on its
dedicated cores, which MWAIT-idle between generations --- exactly
the ``dedicated inference cores'' deployment. Because both modes
are the same binary and the same engine, differences measure
scheduling policy, not implementation drift.

\subsection{Measurement harness and metric integrity}
\label{sec:harness}

\texttt{sched\_bench} (2{,}567 lines) is a self-driving, in-kernel
harness: an autorun matrix (21 cells in the 135M campaign; 17 in
the 7B campaign, whose windows are 25$\times$ longer per cell)
executes on boot with no keyboard, emitting grep-able
\texttt{[sched:\dots] k=v} lines to the NVMe-persisted boot log,
with per-scenario flushes so a late wedge cannot destroy earlier
scenarios. Instrumentation is
armed-gated (one relaxed load when disarmed) and transition-only: a
TSC event ring records \texttt{begin}/\texttt{ack}/\texttt{retract}%
/\texttt{end}/\texttt{migrate} events, and per-core role-time
accumulators update at role transitions --- nothing is added inside
the per-operator barrier loop. Repetitions are interleaved and
report min/median/max with spread; MPERF/APERF ratios are sampled
per cell to expose thermal or frequency drift.

\paragraph{A migration-robust throughput metric (methodology
honesty).} My first general-throughput metric summed the
scheduler's per-process \texttt{delivered} virtual-time counters.
This metric is \emph{contaminated by migration}: on re-enqueue, a
migrated process's virtual time is re-based against the destination
queue's floor (the same seeding CFS-style schedulers perform), so
under heavy lending the summed ``delivered'' rate showed physically
impossible inflation (run-to-run spreads of 118--180\% at lending
duty points, versus 0.1--0.9\% for the corrected metric; one rep
reported more ``work'' than the no-inference ceiling). The headline
metric is therefore \texttt{general\_served\_s}: a monotonic
per-slot \emph{served-quanta} counter incremented once per quantum
actually executed, never re-based, summed across processes. At zero
migration the two metrics agree exactly (all static cells and the
duty-0 cell report identical values for both); under migration only
the served counter is meaningful. I report this explicitly because
the naive metric \emph{overstates} elastic's win --- the corrected
numbers below are the honest ones.

\section{Evaluation}
\label{sec:eval}

\subsection{Setup}
\label{sec:setup}

\textbf{Hardware.} AMD Ryzen~9~9800X3D (Zen~5, 8~physical cores,
16~hardware threads, invariant TSC at 4.5\,GHz), DDR5-6000 dual
channel ($\sim$75\,GB/s). Bare metal --- no hypervisor, no host OS.
All numbers are from single-boot autorun campaigns persisted to the
NVMe boot log; MPERF/APERF ratios stayed at 1.10--1.12 across all
cells (no thermal or frequency confound).

\textbf{Models.} Two operating points: SmolLM2-135M (Q4\_0,
87.6\,MB loaded) is the aggressive one --- at
$\sim$0.75\,ms/token it exercises per-token elasticity
$\sim$100$\times$ harder than the large model --- and
Qwen2.5-7B-Instruct (Q4\_0, 4{,}238\,MB loaded) is the
bandwidth-bound one. The work-conservation matrix
(\S\ref{sec:eval-m3}) and the latency and migration
microbenchmarks (\S\ref{sec:eval-m1}) use the 135M point; the
saturation-knee and churn-safety experiments
(\S\ref{sec:eval-m7}, \S\ref{sec:eval-m4}) are reported for both
models. (In the boot-log artifact these are the harness scenarios
\texttt{m3}; \texttt{borrow-lat} and \texttt{m2}; \texttt{m7};
and \texttt{m4}, respectively. The lending-path bit-exactness
check \texttt{bitexact-lend} (\S\ref{sec:eval-m4}) and the
governance-share sweep (\S\ref{sec:eval-c32}) are from a third
single-boot campaign on the same machine.)

\textbf{Workload.} 24 synthetic preemptive tenants --- instances of
the kernel's embedded AOT spin agent, a pure-compute loop with no
syscalls, preempted and migrated by the ordinary scheduler paths
--- deliberately oversubscribing the machine and exercising the
dynamic slot table; one timer-paced latency tenant (2\,ms period,
pinned and non-migratable --- a \emph{disclosed worst case},
\S\ref{sec:eval-m3}); and a generation driver issuing 32-token
generations through the agent syscall path at duty cycle
$d \in \{0,25,50,75,100\}\%$. Achieved duty tracks the target
within 1.4\,pp on every elastic cell (worst single rep $+1.4$\,pp
at duty 25); the worst cell overall is $+2.2$\,pp (static-8 at
duty 50). Windows are 10\,s, 5 repetitions, medians reported;
elastic and static cells are interleaved.

\subsection{Work conservation vs.\ static partitions (headline)}
\label{sec:eval-m3}

\begin{table*}[t]
\centering\small
\begin{tabular}{@{}r rr rr rr@{}}
\toprule
& \multicolumn{2}{c}{\textbf{Elastic}} & \multicolumn{2}{c}{\textbf{Static-8}} & \multicolumn{2}{c}{\textbf{Static-12}}\\
\cmidrule(lr){2-3}\cmidrule(lr){4-5}\cmidrule(lr){6-7}
\textbf{Duty} & general/s & infer fwd/s & general/s & infer fwd/s & general/s & infer fwd/s\\
\midrule
0\%   & \textbf{653.9} & ---   & 327$^{\dagger}$ & --- & 140$^{\dagger}$ & ---\\
25\%  & \textbf{574.0} & 140.0 & 326.9 & 59.3  & 140.1 & 154.2\\
50\%  & \textbf{496.8} & 275.5 & 327.0 & 120.8 & 140.3 & 305.7\\
75\%  & \textbf{417.1} & 418.8 & 326.9 & 177.9 & 140.2 & 456.3\\
100\% & 326.9          & 571.7 & 327.0 & 232.4 & 140.2 & 601.3\\
\bottomrule
\end{tabular}
\caption{\textbf{Work conservation versus static partitions.}
General throughput (\texttt{general\_served\_s}, migration-robust
served-quanta/s, median of 5) and inference throughput (forward
passes/s) versus inference duty cycle; 135M model, 24 tenants,
10\,s windows, single boot. Run-to-run spread of the headline metric is
0.1--0.9\% in every cell. $^{\dagger}$Static general throughput is
duty-independent by construction (placement never uses the
inference cores regardless of duty; measured flat to $<$0.2\%
across duty 25--100\%), so the duty-0 value equals the measured
flat line.}
\label{tab:m3}
\end{table*}

Table~\ref{tab:m3} is the paper's central result. Because both
modes are the same shipped binary (\S\ref{sec:staticmode}), the
sweep is an ablation of scheduling policy alone. Reading the
general-throughput column: the elastic curve is strictly monotone
in duty --- 653.9, 574.0, 496.8, 417.1, 326.9 served-quanta/s ---
while static-8 is flat at $\approx$327 and static-12 flat at
$\approx$140. Three observations:

\textbf{(1) At intermediate duty --- the region a deployment
actually lives in --- elastic strictly dominates static-8 on both
axes.} At duty 25/50/75\% elastic delivers
1.75$\times$/1.52$\times$/1.28$\times$ static-8's general
throughput (574.0/496.8/417.1 vs.\ $\approx$327) while
simultaneously delivering more inference throughput. The endpoints
behave as the design predicts: at duty 0\% elastic recovers
\emph{all eight} cores a static split leaves idle (653.9 vs.\
327 --- expected of any work-conserving design, but measured, and
static deployments really do strand it), and at duty 100\% the
general axes converge (326.9 vs.\ 327.0). There is no operating
point at which static-8 wins anything. (Static-8's
\emph{inference} axis under this matrix is additionally depressed
by SMT-sibling contention from the 24 crammed tenants --- an
effect I analyze separately below and in
\S\ref{sec:limitations}, and deliberately do not headline.)

\textbf{(2) No fixed $K$ tracks the frontier.} Static-12 buys the
best dedicated-inference throughput (601.3 fwd/s at duty 100) at
the price of stranding general throughput \emph{everywhere}:
elastic delivers 2.3--4.7$\times$ static-12's general throughput
across the sweep while reaching 90--95\% of its inference
throughput. Choosing $K$ requires knowing the duty cycle in
advance, and any choice is wrong at every other duty point; the
elastic gang auto-sizes online, per generation, with no
configuration.

\textbf{(3) Lending really fired.} The elastic cells performed up
to 301 lend-migrations per 10\,s window (91/189/294 at the
duty-25/50/75 median reps), i.e.\ the displaced-tenant machinery
--- park, migrate, rebalance, return --- is on the hot path of
these numbers, not a theoretical capability. The monotone,
low-spread elastic curve \emph{is} the aggregate evidence that
displaced tenants keep running: the general population's
throughput degrades exactly in proportion to the cycles genuinely
consumed by inference, not by lending mechanics.

\paragraph{The inference column, honestly.} Elastic's inference
throughput exceeds static-8's at every duty point (e.g.\ 571.7
vs.\ 232.4 fwd/s at duty 100), but this gap is \emph{not} a lending
effect --- at duty 100 both run the gang at native width 8 and
lending is essentially quiescent (0--7 lend-migrations per window).
It is a placement/SMT effect: the static partition crams 24 tenants
onto the 8 hardware threads that are SMT siblings of its dedicated
inference threads, so the ``dedicated'' partition's inference
suffers collateral port and bandwidth contention that elastic
placement avoids by spreading load across the whole machine. I
report the numbers as measured, attribute them to placement rather
than to the membership protocol, and flag the SMT topology as a
confound in \S\ref{sec:limitations}. The paper's dominance claims
rest on the general-throughput axis, where the comparison is
unambiguous.

\paragraph{Disclosed worst case: the pinned latency tenant.} The
latency tenant is deliberately pathological: pinned to a lendable
core and non-migratable, so when its core is lent it can only wait.
Its p99 completion latency is $\approx$21\,ms at duty 0 (pure
queueing under 24-tenant oversubscription), 175--183\,ms at duty
25--75\% (bounded by one 32-token generation plus queueing, as the
protocol predicts --- the tail is one generation because return is
prompt), and 2.8\,s at duty 100\% (back-to-back generations; a
pinned tenant on a lent core starves for as long as the burst
train lasts). A migratable tenant --- the common case --- pays only
the migration path (\S\ref{sec:eval-m1}). I include the pinned
case because it is the honest upper bound of what lending can do
to a process that refuses to move.

\subsubsection{Baseline fairness gates}
\label{sec:eval-sanity}

A static baseline is only meaningful if it is \emph{good at its
job}. Two sanity gates, run in the same boot: (i) static-$K$'s
inference throughput must approach a solo run at width $K$ (no
tenants at all); (ii) its general throughput must approach a
no-inference run on its $16{-}K$ cores. Measured (8-tenant
moderate load): static-8 achieves \textbf{93.3\%} of solo
inference (587.4 vs.\ 629.4 fwd/s) and \textbf{99.9\%} of
no-inference general throughput; static-12 achieves
\textbf{99.5\%} and \textbf{100.0\%}. The baselines are not
strawmen --- each is within a few percent of the best any static
split of its size could do.

\subsection{An external anchor: Linux EEVDF on the same silicon}
\label{sec:eval-linux}

The static-$K$ ablation isolates policy inside one binary; a
reviewer will still ask what Linux plus \texttt{taskset} does on
this machine. I therefore ran llama.cpp~\cite{llamacpp} decode
under Ubuntu (kernel 6.17.0-35-generic,
EEVDF/\texttt{SCHED\_OTHER}, performance governor) on the same
9800X3D, against a general load of 24 userspace busy-counter
processes. This is a \emph{directional anchor, not an
apples-to-apples port}: the engine and the OS both differ, so I
compare shapes and ratios, never absolute units against
Anima~OS's served-quanta/s.

\begin{table}[t]
\centering\small
\begin{tabular}{@{}l rr@{}}
\toprule
\textbf{llama.cpp decode, 8 threads} & \textbf{135M} & \textbf{7B}\\
\midrule
unpinned, no load (tok/s)   & 1230 & 14.63\\
unpinned $+$ load (tok/s)   & 1.5  & 0.235\\
\quad collapse              & $\sim$800$\times$ & $\sim$62$\times$\\
pinned (\texttt{taskset}) $+$ load (tok/s) & $\sim$975$^{*}$ & ---\\
\bottomrule
\end{tabular}
\caption{\textbf{External Linux baseline} (same machine; Ubuntu
kernel 6.17.0-35-generic, EEVDF, performance governor; ``load''
$=$ 24 busy-loop processes; medians, $n{=}3$ per cell). A
directional anchor only --- different engine and OS.
$^{*}$Fragile: one of three pinned runs still collapsed to
0.95\,tok/s; at 16 threads, every unpinned$+$load 135M run timed
out.}
\label{tab:linux}
\end{table}

Table~\ref{tab:linux} tells a three-way story. \textbf{(1) Linux
unpinned: the gang collapses.} Under load, unpinned 8-thread
decode falls to 1.5\,tok/s on the 135M model (individual runs
1.01 and 1.92; a third timed out) --- an $\sim$800$\times$
collapse from 1230\,tok/s unloaded --- and to 0.235\,tok/s on the
7B ($\sim$62$\times$ from 14.63). At 16 threads unpinned under
load, every 135M run timed out. This is the textbook gang
pathology of \S\ref{sec:notfungible}, measured on identical
silicon: EEVDF preempts gang threads independently, and every
barrier stalls on whichever participant is descheduled.
\textbf{(2) Linux static pin: protects inference, strands the
partition.} Pinning the gang threads restores $\sim$975\,tok/s
under load --- but the cell is fragile (one of three runs still
collapsed to 0.95\,tok/s), and the pin is exactly Anima~OS's
static-$K$ baseline: in a duty-cycle sweep of the pinned
configuration, the general population's throughput is capped at
$\sim$4.14 core-equivalents even at inference duty 0 (inference
entirely idle) --- roughly half the 8-core budget stranded
whenever inference is not running. \textbf{(3) The elastic gang
gets both.} Gang-protected --- bit-exact under real membership
change (\S\ref{sec:eval-m4}) --- \emph{and} work-conserving: the
duty-0 cell of Table~\ref{tab:m3} recovers every stranded core.
The robust external claim is the unpinned collapse, consistent
across runs and models; the pinned-under-load and 16-thread cells
are noisy and timeout-prone at small $n$, and I weight them
accordingly.

\subsection{The saturation knee: why elasticity is nearly free}
\label{sec:eval-m7}

\begin{table}[t]
\centering\small
\begin{tabular}{@{}r rr@{}}
\toprule
\textbf{Gang width} & \textbf{7B tok/s} & \textbf{135M tok/s}\\
\midrule
1  & 4.36  & 270.4\\
2  & 7.50  & 479.1\\
4  & 10.72 & 856.0\\
8  & \textbf{12.72} & \textbf{1352.5}\\
12 & 12.76 & 1272.1\\
16 & 12.83 & 1277.5\\
\bottomrule
\end{tabular}
\caption{\textbf{Solo decode throughput vs.\ gang width}
(median of 3; spreads 0.0--3.7\%). Both models saturate at width
8 ($=$ the machine's physical core count); the 7B model gains
$<$1\% from widths 9--16, and the 135M model \emph{loses}
$\approx$6\%.}
\label{tab:m7}
\end{table}

Table~\ref{tab:m7} answers the two stock objections at once.
\emph{``Why not dedicated cores?''} --- because past width 8 they
buy nothing, and this holds in \emph{two different memory
regimes}. The 7B model is pinned to the DRAM bandwidth floor
(78.6\,ms/token at width 8, against a $\sim$55\,ms theoretical
floor for a 4.2\,GB weight stream at 75\,GB/s). The 135M model is
\emph{not} DRAM-bound: its 87.6\,MB weight set fits in the
9800X3D's 96\,MB stacked L3, and its width-8 throughput implies a
$\sim$118\,GB/s weight-streaming rate --- above the 75\,GB/s DRAM
ceiling, i.e.\ it decodes out of cache. Its knee at 8 is therefore
core-count/SMT saturation, not bandwidth, and past it barrier and
SMT-sibling overhead exceed the marginal compute, so throughput
declines. Either way the conclusion is the same: cores 9--16 are,
from the gang's perspective, nearly free to cede --- which is
precisely the population the elastic scheduler lends back to
general work --- and the mechanism is profitable in both the
DRAM-bound and the cache-resident regime.
(\S\ref{sec:eval-linux} prices the dedicated-cores alternative
externally: a Linux static pin strands about half its core budget
whenever inference is idle.)
\emph{``Why not a GPU?''} --- the target class is GPU-less edge
and appliance deployments, and the trade structure is intrinsic to
any engine that saturates below machine width:
over-provisioning the gang hurts,
under-provisioning the general population hurts, and only an
elastic membership can sit at the knee of both simultaneously. The
knee also explains \emph{why} Table~\ref{tab:m3}'s elastic curve
can approach static-12's inference throughput while doubling and
quadrupling its general throughput: the marginal inference value
of the lent cores is small by physics.

\subsection{Safety: bit-exactness under membership churn}
\label{sec:eval-m4}

Work conservation would be worthless if elastic membership could
corrupt a barriered computation (\S\ref{sec:notfungible}); recall
that in this kernel, logits feed the safety
gate~\cite{animaos-mcp}, so a lost or extra gang core is a
correctness event, not a performance event. The churn experiment
compares outputs byte-for-byte against a fixed-width reference
run: same prompt, greedy sampling, 16 tokens; four back-to-back
generations with target widths $\{8,16,12,15\}$.

Being precise about what membership schedule this exercises: the
width \emph{target} changes per generation, and within each
generation cores \emph{join} at token boundaries as their acks
land during the ramp (the per-token latch admitting them one token
at a time --- I1). Mid-token \emph{departure} does not occur ---
not because the test declines to force it, but because the
protocol makes it unrepresentable: departure happens only via an
epoch bump, and epoch bumps happen only at generation boundaries
(\S\ref{sec:ack}). That is the design working as intended --- the
deadlock of \S\ref{sec:notfungible}(1) is avoided by
construction rather than survived --- and the schedule tested here
(per-token joins, generation-boundary departures) is exactly the
schedule the protocol can produce in deployment.

On the 135M model the participant mask changed 8 times through 6
distinct participant sets across the churn run (versus 1 mask in
the reference --- the \texttt{churn\_effective} gate proves the
test actually exercised membership change, not a no-op knob); on
the 7B model, 6 changes through 4 distinct sets. In both cases the
token byte stream and the final-logits hash are
\textbf{byte-identical} to the fixed-width reference
(first-divergence index $-1$), and a deliberately corrupted
negative control (\texttt{neg\_control\_detected}) confirms the
equality test is capable of failing. The mechanism-level reason is
the one the design section argues: matvec rows are
value-independent of \emph{which} worker computes them, and the
latch guarantees each row of each wave is computed exactly once by
exactly one member of that token's snapshot.

\paragraph{Closing the width-vs-lending gap:
\texttt{bitexact-lend}.} The churn schedule above drives
membership through the width \emph{target}, so on its own it
proves width-invariance; a reviewer may fairly ask whether the
\emph{actual lending path} is equally safe. A second experiment
(\texttt{bitexact-lend}; 135M, silicon, from a follow-up
single-boot campaign) pins the width knob completely --- gang
width fixed at 8, zero width flips; the width machinery is never
touched --- and lets \emph{real lending} drive per-token
membership against 24 tenants (2 generations, 16 tokens each).
Lending genuinely fired: 14 lend events and 3 retractions
produced 5 ack-driven participant-mask changes, including 3
participant sets that never occur in a no-lending reference run;
a \texttt{lending\_effective} gate (lend events $>0$ $\wedge$
new-vs-reference masks $>0$ $\wedge$ width unchanged) certifies
that the \emph{only} membership driver was the ACK latch. The
output is byte-identical to the no-lending reference: the same
73-byte token stream (first-divergence index $-1$) and the same
final-logits hash (\texttt{0x9a2917eca1a62bd1}), with the
deliberately corrupted negative control again detected. Together
the two experiments cover both membership drivers the protocol
has --- forced width churn, and genuine ack-latch lending with
the width pinned --- and the output is bit-exact under each.

\subsection{Borrow and return latency}
\label{sec:eval-m1}

Three costs matter, and they live on three different timescales; I
report them separately rather than blending them into one
misleading number.

\textbf{(a) The per-token participation latch is near-free.}
Deciding membership for a token is one tagged atomic read per
candidate core inside \texttt{participant\_mask}
(Listing~\ref{lst:latch}) --- this is \emph{why} per-token
membership change is possible at all, and its correctness (not
its cost) is what \S\ref{sec:eval-m4} verifies. There is no
per-token handshake, no IPI, and no lock.

\textbf{(b) Returning a core is sub-microsecond.} From the
\texttt{gang\_end} epoch bump to a lent core's ack retraction:
p50 $=0.22\,\mu$s, p99 $=0.38\,\mu$s, max $=0.42\,\mu$s
($n{=}224$ transitions over 32 forced borrow/return cycles across
7 cores). The doorbell-wake path out of MWAIT plus one epoch
comparison --- the core is back in its scheduler loop picking a
tenant within a microsecond of the generation ending.

\textbf{(c) Acquiring a \emph{busy} core costs one scheduling
quantum --- by design.} With all lendable cores running tenants
(24-tenant oversubscription), the time from \texttt{gang\_begin}
to a core's ack is one WRR quantum: a running tenant is
deliberately \emph{not} preempted mid-slice for lending, so the
lent core acks when its current quantum expires. The measurements
($n{=}224$ over 32 forced borrow/return cycles, 7 cores) say
exactly that and nothing more: per-core medians of
15.2--16.4\,ms --- the spread across cores reflecting staggered
quantum phases, with within-core variance near zero --- and a
worst observed acquisition of 20.1\,ms (a quantum plus scheduler
pass slack). I deliberately do not quote tail percentiles here:
this distribution is a deterministic phase, not a tail. The cost
is amortized over the burst --- one quantum buys the core for the
whole generation train --- and the never-wait property means the
generation does not stall while waiting: it runs at the
already-acked width and widens as acks arrive. It would be easy to
quote the $\mu$s-scale numbers (a,b) as ``reconfiguration
latency''; the honest summary is: \emph{return and per-token
re-membership are microsecond-scale; acquisition of a busy core is
quantum-scale by policy}.

\textbf{(d) Migration mechanics.} Measured in a dedicated
displaced-tenant burst experiment (a single 118\,ms, 48-token
generation against 24 tenants; 7 cores lent at peak, 7 tenants
migrated, with per-tenant progress traces confirming the migrated
tenants kept advancing on their new cores): the migration
primitive (CAS \texttt{RUNNABLE}$\to$\texttt{MIGRATING} plus
re-enqueue) costs p50 $=0.22\,\mu$s, max $=0.7\,\mu$s per
migration. (That this p50 numerically equals the return-path p50
of (b) is coincidence, not a copy error --- they are different
code paths, instrumented separately, landing on the same
sub-microsecond scale.) Migrations are paced at $\le$2 per
scheduler pass --- a deliberate service-latency bound, so a
fully-lent machine drains its displaced tenants over a few passes
rather than in one burst. Within the work-conservation windows of
\S\ref{sec:eval-m3} this machinery ran up to 301 times per 10\,s
without perturbing the 0.1--0.9\% headline spreads.

\subsection{Governance-modulated inference share}
\label{sec:eval-c32}

The kernel maintains a per-agent constitutional trust score, and
the general scheduler already consumes trust as a WRR weight. The
elastic gang now consumes it too: trust caps how many general
cores the gang may \emph{borrow} --- a \textbf{lend-admission
quota}, not a literal per-token CPU budget, because a core cannot
be safely withdrawn mid-generation (it would stall the barrier).
The quota is a pure admission conjunct evaluated \emph{before} a
core acks, so a quota-withheld core is indistinguishable from a
not-yet-acked core --- no barrier can hang on it
(\S\ref{sec:latch}), and work-stealing absorbs its rows. Weights
are clamped to $[1,4]$ relative to the registry's maximum live
trust, so uniform trust means full quota --- identical to the
shipped full-lend behavior (the elastic cells of
Table~\ref{tab:m3} are the trust-4 point of this dial).

\begin{table}[t]
\centering\small
\begin{tabular}{@{}c c c r r@{}}
\toprule
\textbf{Trust} & \textbf{Quota} & \textbf{Borrowed} & \textbf{infer fwd/s} & \textbf{general/s}\\
\midrule
1 & 1/4 & 2 & 343.82 & 559.89\\
2 & 2/4 & 4 & 460.76 & 467.58\\
3 & 3/4 & 6 & 510.20 & 374.04\\
4 & 4/4 & 7 & 572.35 & 327.36\\
\bottomrule
\end{tabular}
\caption{\textbf{Governance-modulated inference share}:
trust-weighted lend-admission quota, on silicon. Cores borrowed,
inference throughput (forward passes/s), and general throughput
(\texttt{general\_served\_s}, served-quanta/s) versus the owner
agent's trust weight; 135M model, 24 tenants, 8\,000\,ms windows,
32 tokens/burst. Trust 4 (full quota) reproduces the shipped
full-lend behavior (general $\approx$327, Table~\ref{tab:m3}).}
\label{tab:c32}
\end{table}

Table~\ref{tab:c32} sweeps the owner's trust weight (via a bench
override; 135M model, 24 tenants, 8\,000\,ms windows, 32
tokens/burst). Both curve gates emitted true on silicon
(\texttt{share\_rises\_with\_trust},
\texttt{general\_falls\_with\_trust}): inference throughput rises
monotonically by 66\% across the trust range
(343.82$\to$572.35\,fwd/s) while general throughput falls 42\%
(559.89$\to$327.36 served-quanta/s). The internal consistency is
the point: at trust 4 (full quota) the general axis lands at
327.36 --- the static-8/duty-100 endpoint of Table~\ref{tab:m3}
($\approx$327), i.e.\ full lend \emph{is} today's behavior ---
and at trust 1 it recovers toward the duty-0 end (559.89, heading
toward 653.9). Governance-modulated share is therefore not a new
mechanism but a \emph{dial between the elastic sweep's
endpoints}: the constitutional trust score acts as a first-class
scheduler lever over the gang's core share.

\subsection{When to use the elastic gang --- and when not}
\label{sec:envelope}

The measurements above delimit an applicability envelope, and it
is worth stating plainly.

\textbf{The mechanism pays off when three conditions hold.}
(i)~\emph{Decode saturates below machine width.} The knee of
Table~\ref{tab:m7} --- width 8 on this 16-thread machine, in both
the DRAM-bound and the cache-resident regime --- is what makes
lent cores nearly free for inference; elasticity is profitable
\emph{because} the marginal inference value of the cores it lends
away is small by physics (\S\ref{sec:eval-m7}).
(ii)~\emph{Inference is bursty, or its duty cycle is unknown in
advance.} The entire gap between the elastic curve and every
static split in Table~\ref{tab:m3} lives in the duty-varying
region: at duty 25--75\% elastic delivers 1.28--1.75$\times$
static-8's general throughput at equal-or-better inference, and
at idle it recovers all eight stranded cores. Choosing a static
$K$ requires forecasting the duty cycle; the elastic gang needs
no forecast. (iii)~\emph{Displaced work is migratable.} The
park--migrate--resume path is what keeps the general population
running under lending (\S\ref{sec:eval-m1}d); latency-critical
work should either remain migratable or be pinned to the
non-lendable tier the mechanism already provides.

\textbf{At sustained saturation the mechanism buys nothing ---
and costs nothing.} At duty 100\% the elastic general axis
converges to static-8 (326.9 vs.\ 327.0 served-quanta/s,
Table~\ref{tab:m3}), and lending is essentially quiescent. A
deployment that decodes back-to-back around the clock can pin a
static partition at the knee and match elastic's general
throughput; elasticity's remaining value at that endpoint is
only that it \emph{found} the knee online. (The inference-axis
gap that persists at duty 100 is a placement/SMT effect, not a
lending effect --- \S\ref{sec:eval-m3},
\S\ref{sec:limitations}.)

\textbf{Where it does not help.} First, if decode does \emph{not}
saturate below machine width --- a machine narrower than the
engine's knee, or an engine whose scaling continues to the last
core --- then every lent core costs inference throughput
directly, and lending becomes a genuine capacity split rather
than the nearly-free cession measured here. The knee is the
precondition; it should be measured (as in \S\ref{sec:eval-m7})
before lending is enabled. Second, a workload dominated by
pinned, non-migratable processes on lendable cores inherits the
disclosed worst case --- p99 completion 2.8\,s at duty 100\%
(\S\ref{sec:eval-m3}) --- rather than the microsecond-scale
common case. Third, gang ramp-up is quantum-scale by policy
(\S\ref{sec:eval-m1}c): a busy core joins only when its tenant's
quantum expires, so an isolated generation much shorter than a
quantum on a fully busy machine runs mostly at its already-acked
width; the acquisition cost amortizes only over generation
\emph{trains}. Finally, the target class is CPU decode on
GPU-less edge and appliance machines (\S\ref{sec:eval-m7});
where inference runs on a dedicated accelerator, the CPU gang
this paper elasticizes does not exist in the first place.

\section{Related Work}
\label{sec:related}

I position along four axes: \textbf{A1} a real kernel owns the
scheduler (not a userspace runtime or a Linux module);
\textbf{A2} a hard-barriered parallel computation is a first-class
kernel-schedulable entity \emph{with runtime-mutable membership};
\textbf{A3} work-conserving per-work-item core borrow/return where
displaced \emph{processes} migrate and keep running; \textbf{A4}
co-scheduling against a general preemptive OS process population.
Each family below misses at least one.

\begin{table}[t]
\centering\scriptsize
\setlength{\tabcolsep}{4pt}
\begin{tabular}{@{}lcccc@{}}
\toprule
System family & A1 & A2 & A3 & A4\\
\midrule
Classic gang sched.~\cite{ousterhout82,feitelson92} & $\sim$ & $\sim$ & -- & $\sim$\\
DBC / FCS / Paired~\cite{sobalvarro95,fcs03,wiseman03} & $\sim$ & $\sim$ & -- & $\sim$\\
Runtime co-sched.~\cite{tucker89,mccann93,schedact,lithe,callisto} & -- & $\sim$ & $\sim$ & $\sim$\\
Shenango / Caladan / IX~\cite{shenango,caladan,ix} & -- & -- & $\sim$ & $\checkmark$\\
VMware co-scheduling~\cite{vmware-cosched} & $\checkmark$ & $\sim$ & -- & $\sim$\\
GPU serving~\cite{orca,vllm,sarathi,muxserve} & -- & $\sim$ & -- & --\\
AIOS~\cite{aios-colm} & -- & $\sim$ & -- & --\\
coconutOS~\cite{coconutos} & $\checkmark$ & -- & -- & $\sim$\\
Engine threading~\cite{llamacpp} & -- & $\sim$ & -- & --\\
\textbf{Elastic gang (this work)} & $\checkmark$ & $\checkmark$ & $\checkmark$ & $\checkmark$\\
\bottomrule
\end{tabular}
\caption{Positioning. $\checkmark$ = holds the axis; $\sim$ =
partially; -- = does not. A2 is deliberately stated as a neutral
property (kernel-schedulable barriered entity with runtime-mutable
membership) so that gang schedulers and vCPU co-scheduling can
score partial credit.}
\label{tab:related}
\end{table}

\textbf{Classic gang and co-scheduling.} Ousterhout introduced
gang scheduling for fine-grain synchronized
programs~\cite{ousterhout82}; Feitelson and Rudolph quantified its
benefit~\cite{feitelson92}. The refinement line relaxed
\emph{which} jobs must be co-resident: demand-based coscheduling
schedules communicating processes together on
demand~\cite{sobalvarro95}; Flexible CoScheduling classifies
processes by communication behavior and co-schedules only those
that need it~\cite{fcs03}; Paired Gang Scheduling pairs
compute-bound with I/O-bound gangs on the same
slots~\cite{wiseman03}. All operate at job/timeslice granularity
on (typically dedicated) cluster nodes, and none changes the
\emph{membership of a running barrier group}: the elastic gang's
per-token membership change under a hard barrier
(\S\ref{sec:notfungible}) is precisely the case this literature
brackets out.

\textbf{Cooperative core-sharing between parallel runtimes.} A
second family makes parallel computations \emph{malleable} so that
runtimes can share cores. Tucker and Gupta's process control
shrinks each application's worker count to match its processor
allocation, with workers suspending at safe
points~\cite{tucker89}; McCann et al.\ studied dynamic processor
(re)allocation policies across such malleable jobs~\cite{mccann93};
scheduler activations give a user-level thread system kernel
upcalls on processor allocation and revocation~\cite{schedact};
Lithe composes parallel libraries by handing \emph{harts} between
cooperative schedulers~\cite{lithe}; and Callisto co-schedules
multiple parallel runtime systems over a shared core pool,
absorbing core loss via fine-grained work items~\cite{callisto}.
This family genuinely anticipates the \emph{spirit} of lending;
four differences separate this work from it. (i)~Their
computations cede cores at \emph{application-chosen cooperative
safe points} --- loop-chunk boundaries, scheduler upcalls,
work-item boundaries. The elastic gang's computation has no such
freedom: every token is a hard barrier chain in which a latched
core is \emph{obligated}, and safety must be established
protocol-side --- the formal never-wait-on-a-named-core invariant
of \S\ref{sec:latch} --- not by asking the computation to yield.
(ii)~They are userspace runtimes over pthreads on a commodity
kernel (or, for scheduler activations, a kernel interface serving
user-level threading); the OS scheduler underneath is not theirs
(A1). (iii)~The counterpart they share cores with is another
parallel runtime's \emph{threads}: none co-schedules against the
OS's own general preemptive process population with kernel-side
park-and-migrate of displaced processes (A3/A4). (iv)~None proves
output bit-exactness under membership change on silicon --- for
them a lost core is a performance event; here it is a correctness
and safety event, because the logits feed the kernel's safety
gate~\cite{animaos-mcp}.

\textbf{Microsecond-scale core reallocation.} Shenango reallocates
cores between latency-critical and batch applications on a
5-$\mu$s IOKernel interval~\cite{shenango}; Caladan extends the
line, detecting and reacting to memory-subsystem interference at
microsecond timescales~\cite{caladan}; IX contributed the
protected-dataplane architecture (with adaptive batching) that
this line builds on~\cite{ix}. This is the closest mechanism
family in actuation speed, and a
reviewer will reach for Caladan first --- the contrast is
therefore worth stating exactly. First, both are Linux-hosted
(kernel module plus userspace runtimes), so the reallocated cores
run \emph{their} threads, not the OS's general process population
(A1/A4). Second --- the deeper difference --- their batch workload
is \emph{fungible threads}: a revoked core's work is picked up by
any other worker, so revocation needs no coordination with the
computation's structure. My borrowed cores join a hard-barriered
SIMD computation in which an unannounced departure deadlocks a
barrier and an unannounced arrival silently corrupts logits
(\S\ref{sec:notfungible}); the ACK-latch/epoch/gen-tag machinery
exists precisely because the gang is \emph{not} fungible. Caladan
solves ``which app gets the core''; this paper solves ``how a
barriered computation can survive the core coming and going.''

\textbf{Hypervisor co-scheduling.} VMware's relaxed co-scheduling
manages skew between the vCPUs of an SMP VM, stopping and starting
vCPUs to bound divergence~\cite{vmware-cosched}. It shares the
barrier-aware motivation but neither lends vCPUs across
populations per-work-item nor exposes membership change to the
computation.

\textbf{Work-conserving reclamation and colocation QoS.}
``Work-conserving'' has an established real-time reading: GRUB
greedily reclaims the unused bandwidth of constant-bandwidth
servers~\cite{grub00}, and Linux's \texttt{SCHED\_DEADLINE}
carries this reclaiming into a production
kernel~\cite{sched-deadline} --- but what is reclaimed there is
CPU \emph{bandwidth} among sequential real-time tasks, not the
membership of a running barriered gang. At datacenter scale,
Heracles~\cite{heracles} and PARTIES~\cite{parties} reclaim
cores, cache, and bandwidth for batch work under a
latency-critical QoS constraint --- userspace controllers over
Linux, actuating on second-plus feedback intervals, again over
fungible threads. Both lines share this paper's
strand-no-resources goal; neither faces its mechanism problem.

\textbf{GPU LLM serving.} Orca's iteration-level
scheduling~\cite{orca}, vLLM's PagedAttention~\cite{vllm},
Sarathi-Serve's chunked prefills~\cite{sarathi}, and MuxServe's
spatial-temporal GPU multiplexing~\cite{muxserve} all schedule at
token/iteration granularity --- a family resemblance to per-token
lending --- but inside a dedicated accelerator, in userspace, over
\emph{requests}, not OS processes: the resource being multiplexed
is never contended by a general-purpose preemptive population
(A1/A3/A4). The elastic gang is, in effect, iteration-level
scheduling applied to the host CPU itself, where the ``other
tenant'' is the operating system's own workload.

\textbf{AI-native OS work.} AIOS builds agent-serving abstractions
(scheduling, memory, tool calls) as a userspace layer over
Linux~\cite{aios-colm} --- complementary goals, but the CPU
scheduler underneath is stock Linux (A1/A3). coconutOS is a
bare-metal Rust microkernel for AI inference~\cite{coconutos},
but targets GPU partitions with round-robin CPU scheduling and
runs in QEMU; it does not treat a CPU-SIMD barrier gang as a
schedulable entity. Engine-level threading in
llama.cpp~\cite{llamacpp} tunes thread counts \emph{within} the
engine and documents the same bandwidth saturation my width sweep
(\S\ref{sec:eval-m7}) shows, but has no OS co-scheduling: on a general-purpose host the
engine's threads and everyone else's meet in a scheduler that
knows nothing about barriers. Proportional-share
scheduling of the general population follows the classic
virtual-time lineage~\cite{lottery}, and Theseus is prior art for
ambitious single-contributor Rust OS structure
research~\cite{theseus}.

\section{Limitations}
\label{sec:limitations}

\textbf{Single machine, single vendor.} Every number is from one
Zen~5 box. The mechanism is ISA-generic (atomics, MWAIT, TSC), and
the two-model matrix spans a 100$\times$ token-time range, but the
absolute numbers --- the knee at 8, the 16\,ms quantum --- are
machine-specific. The contribution is the mechanism plus its
measured properties, not the constants.

\textbf{SMT.} The 16 ``cores'' are 8 physical cores $\times$ 2
threads. Lending an SMT sibling of a busy core is not lending a
physical core: siblings share execution ports and, critically for
a bandwidth-bound gang, memory bandwidth. This cuts both ways ---
it inflates neither population's axis uniformly --- and it is the
most likely explanation for static-8's inference degradation under
tenant oversubscription (\S\ref{sec:eval-m3}). In particular, the
duty-100 inference gap between elastic and static-8 (571.7 vs.\
232.4 fwd/s) is placement/SMT-dominated, not a lending effect ---
both gangs run at native width 8 and lending is quiescent there
--- which is why that ratio appears nowhere in this paper's
headline claims. A one-thread-per-physical-core pinning study is
future measurement work.

\textbf{The Linux baseline is a directional anchor.} The
static-$K$ baseline isolates the \emph{policy} question inside
one binary (same engine, same kernel), which is the clean
comparison. The external Linux measurement
(\S\ref{sec:eval-linux}) anchors the collapse-vs-strand trade on
the same machine, but with a different engine (llama.cpp) and a
different scheduler (EEVDF), so it supports shape and ratio
claims only --- never unit-for-unit comparison against Anima~OS's
served-quanta/s --- and its pinned-under-load and 16-thread cells
are noisy and timeout-prone at small $n$. A controlled port of
the same engine to a Linux userspace harness remains future work.

\textbf{Busy-core acquisition is quantum-bound, not
microsecond-scale.} By policy (no mid-slice preemption for
lending), grabbing a tenant-occupied core costs $\approx$16\,ms
(\S\ref{sec:eval-m1}c). A preempt-for-lend policy would trade
tenant latency for gang ramp time; I have not explored it. The
per-token latch and the return path are the microsecond-scale
parts.

\textbf{Pinned non-migratable processes.} The disclosed worst case
(\S\ref{sec:eval-m3}): a process pinned to a lendable core and
unable to migrate waits out the burst train --- 2.8\,s p99 at
100\% duty. Deployments should pin latency-critical work to
non-lendable cores (the mechanism already supports a never-lends
tier) or leave it migratable.

\textbf{Metric caveat.} The headline metric replaced a
migration-contaminated one (\S\ref{sec:harness}); while the served
counter is monotone and migration-robust by construction, quanta
are its unit --- it counts scheduling service, which under
identical quantum lengths is proportional to CPU time delivered,
but it is not an application-level operations/s metric.

\textbf{Governance-share sweep scope.} The trust dial
(\S\ref{sec:eval-c32}) is swept via a bench override of the
owner's weight, on one model and one workload matrix, and its
granularity is intrinsically admission-time: once admitted, a
core serves until the generation ends. A study of the quota under
deployment-driven trust dynamics is future work. No headline
claim depends on it, and it is reported as a measured property of
the substrate, not as a separate contribution.

\section{Conclusion}
\label{sec:conclusion}

A hard-barriered LLM inference gang and a general preemptive
process population can share one machine's cores without a static
line between them. The requirements are a
membership protocol that never lets a barrier wait on a named core
--- an epoch-tagged ACK latch, snapshotted per token into a single
generation-tagged word --- and a scheduler that parks, migrates,
and resumes the displaced processes as ordinary work. On real
silicon, inference output remains bit-exact under verified
per-token membership change, cores return to general work within a
microsecond of a generation ending, and the policy ablation shows
what the mechanism buys: 1.75$\times$/1.52$\times$/1.28$\times$
the general throughput of a static 8-core split at intermediate
inference duty at equal-or-better inference throughput, full
recovery of the stranded cores at idle, convergence at saturation,
and dominance over every fixed $K$ in between. Because decode saturates
at a knee well below the machine's width --- in both the DRAM-bound
and the cache-resident regime --- the
cores the gang cedes are nearly free; the elastic gang simply
puts them where they matter, one token at a time.

\section*{Availability}

Anima~OS's measurement harness emits every number in this paper as
grep-able \texttt{[sched:\dots]} lines persisted to the machine's
NVMe boot log; the two campaign logs and the parsing scripts
accompany the artifact. Mechanism, harness, and baselines are the
same kernel binary.


\end{document}